\documentclass[11pt,a4paper]{article}
\usepackage[dvips]{graphicx}
\begin{document}
\thispagestyle{empty}
\renewcommand{\baselinestretch}{1.2}
\small\normalsize
\frenchspacing
\noindent
{\Large \textbf{Entanglement and reduction in two-photon correlations}}\footnote{Also paper-published in Journal of the Optical Society of America B {\bf31} (4), 765-770 (2014).}
\\
\\
{\bf{Arthur Jabs}}
\renewcommand{\baselinestretch}{1}
\small\normalsize
\\
\\
Alumnus, Technical University Berlin. 

\noindent
Vo\ss str. 9, 10117 Berlin, Germany

\noindent
arthur.jabs@alumni.tu-berlin.de
\\
\\
(16 March 2014)
\newcommand{\rmi}{\mathrm{i}}
\vspace{15pt}

\noindent
{\bf{Abstract.}} For a number of basic experiments on two-photon intensity correlations it is pointed out that the results, which are usually explained in terms of the formalism of canonical field quantization, can also be explained in terms of quantum wavepackets, where reduction and entanglement are taken into account.

\vspace{5pt}

\vspace{3pt}

\noindent
\rule{\textwidth}{.3pt} 
\newcommand{\leer}{\hspace*{\fill}}
\newcommand{\lll}{\hspace*{10pt}}
\newcommand{\hsp}{\hspace*{60pt}}
\newcommand{\psit}{\widetilde{\psi}}
\newcommand{\ki}{k_{\rm i}}
\newcommand{\ks}{k_{\rm s}}
\newcommand{\ixi}{x_{\rm i}}
\newcommand{\ixs}{x_{\rm s}}
\newcommand{\psiil}{\psi_{\rm i}^{\rm L}}
\newcommand{\psiis}{\psi_{\rm i}^{\rm S}}
\newcommand{\psiss}{\psi_{\rm s}^{\rm S}}
\newcommand{\psisl}{\psi_{\rm s}^{\rm L}} 
\newcommand{\psiui}{\psi_{\rm i}}
\newcommand{\psius}{\psi_{\rm s}}
\newcommand{\rmf}{\mathrm{f}}
\newcommand{\rmd}{\mathrm{d}} 
\newcommand{\sbl}{\hspace{1pt}}
\newcommand{\bfitp}{\emph{\boldmath $p$}}
\newcommand{\bfitr}{\emph{\boldmath $r$}}
\newcommand{\bfitsr}{\emph{\footnotesize{\boldmath $r$}}}
\newcommand{\bfitQ}{\emph{\boldmath $Q$}}
\newcommand{\bfitk}{\emph{\boldmath $k$}}
\newcommand{\PSI}[1]{\Psi_{\textrm{\footnotesize{#1}}}}
\newcommand{\PHI}[1]{\Phi_{\textrm{\footnotesize{#1}}}}
\newcommand{\spsi}[1]{\psi_{\textrm{\footnotesize{#1}}}}
\newcommand{\pcop}{P_{\textrm{\footnotesize{Cop}}}}
\newcommand{\pred}{P_{\textrm{\footnotesize{red}}}}
\newcommand{\psis}{\psi_{\rm s}(\bfitr,t)}

\noindent
{\textbf{1~~Introduction}}
\smallskip

\noindent
In canonical field quantization the functions that represent the fields and the conjugate field momenta are turned into operators between which commutation relations are postulated, for bosons modeled after the Heisenberg relations and for fermions introduced ad hoc. When the operator functions are expanded in Fourier series, the expansion coefficients are operators, $a,\; a^{\dagger}$, which satisfy commutation relations according to which the product $a^{\dagger}a$ has only non-negative integers as eigenvalues. Here the particle concept enters field theory: the eigenvalues are taken to be particle numbers, $a$ and $a^{\dagger}$ create and destroy the particles in the number or Fock states. In order to avoid that the state with no particles has infinite energy normal ordering, disregarding the commutation relations, is called on. Reduction (collapse) is not mentioned. Canonical field quantization, that here is specifically quantum electrodynamics or quantum optics, correctly describes the experimental findings.

In this note we demonstrate that in a number of experiments the findings can also be correctly described within quantum mechanics, without canonical quantization, when the particles (photons) are introduced in the spirit of Einstein's photon paper [1], with some specifications. These are:

   (1) Photons are not point particles but special wavepackets of electromagnetic waves representing an integral number of \emph{quanta},

   (2) A one-photon wavepacket, when it interacts with a measuring device, is instantaneously contracted to a narrow place (e.g. an atom) and vanishes there: \emph{reduction} [2], [3],
 
   (3) Photon wavepackets may get \emph{entangled}, that is, the wave function representing a system of photons can no longer be written as a product of one-photon wavepackets.

Compared with the machinery of canonical field quantization these conceptions mean a more direct conceptual access to the photon-correlation experiments considered in this note, and, as I think, provide  an intuitively appealing picture. In a number of articles [2 - 5] it has already been shown that the wavepacket approach also avoids many of the conceptual difficulties of nonrelativistic quantum mechanics.

In Section~2 we discuss experiments in which the two entangled signal and idler photon wavepackets (photons, for short) from spontaneous parametric down-conversion only meet and overlap in the measuring device. In Section~3 they overlap at a beam splitter before they enter the measuring device, and in Section~4 they never overlap after they have left the source. Sections~2 and 3 are mainly concerned with reduction, Section~4 with entanglement.

In each experiment, we consider the results obtained when using classical electromagnetic waves; then we take entanglement and reduction into account, and then compare the results with those of quantum optics and experiment. Though basically the radiation is conceived to consist of wavepackets of finite extent, in the mathematical treatment the plane-wave approximation is mostly used.

The most salient difference between classical and quantum predictions appears in the visibility (relative modulation amplitude, intensity correlation function, average correlation product, two-photon detection probability) $V=(C_{\rm max}-C_{\rm min})/(C_{\rm max}+C_{\rm min})$ of the fringes observed in the two-photon intensity correlation as a function of different parameter settings. Therefore our attention is focused on the formulas for this quantity, calculated by quantum optics (qo), quantum mechanics, including reduction and entanglement (qm), and classical optics (co).
\vspace{10pt}

\noindent
{\textbf{2~~Photon overlap in the measuring device}}
\smallskip

\noindent
We begin with the pioneering experiment [6] by Ghosh and Mandel. The joint probability $P$ for the detection of two photons at two points is measured as a function of the separation ($x_1-x_2$) (Fig.~1). Signal and idler photons are produced in
\begin{figure}[h]
\begin{center}
\includegraphics[width=0.70\textwidth]{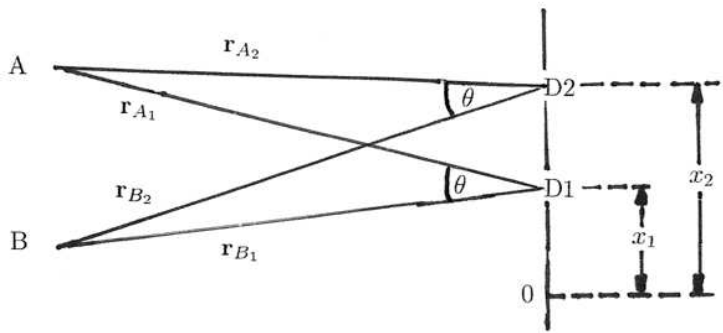}
\caption{Geometry of the interference experiment [6].}
\end{center}
\end{figure}
\noindent
spontaneous parametric downconversion with no definite phase relationships. The intensity of the beams was so low that never more than a single one-photon wavepacket at a time appeared in each beam. 

According to quantum optics the probability $P$ is \{Eq.~(6) in [6]\}
\begin{equation}
P_{\rm qo}(x_1,x_2)\delta x_1\delta x_2=2K_1K_2\delta x_1\delta x_2\{1+\cos(2\pi(x_1-x_2)/L_0)\}.
\end{equation}
Here $\delta x_1$ and $\delta x_2$ are narrow ranges centered at $x_1$ and $x_2$, respecticely. $K_1$ and $K_2$ are scale factors characteristic of the detectors. $L_0\approx\lambda/\theta$ is the spacing of the classical interference fringes when the two waves of wavelength $\lambda$ come together at a small angle $\theta$. Equation~(1) correctly describes the experimental results. The visibility is the factor in front of the cosine, which here is $V_{\rm qo}=1$.

According to classical optics the probability is \{Eq.~(7) in [6]\}:
\begin{equation}
P_{\rm co}(x_1,x_2)\delta x_1\delta x_2=2K_1K_2\delta x_1\delta x_2\left[1+\frac{2\langle|a_A|^2|a_B|^2\rangle}{\langle(|a_A|^2+|a_B|^2)^2\rangle}\cos\frac{2\pi(x_1-x_2)}{L_0}\right].
\end{equation}
Here $a_A$ and $a_B$ are the {\emph c} number coefficients when the classical fields detected at some points $x_1, x_2$ in a distant plane  are written in the form of
\begin{equation}
E(x_1)=a_Ae^{\rmi{\bf k}_{A1}{\bf r}_{A1}}+a_Be^{\rmi{\bf k}_{B1}{\bf r}_{B1}+\rmi\delta},\hspace{10pt}E(x_2)=a_Ae^{\rmi{\bf k}_{A2}{\bf r}_{A2}}+a_Be^{\rmi{\bf k}_{B2}{\bf r}_{B2}+\rmi\delta}
\end{equation}
and averaged over the relative random phase $\delta$ between signal and idler photons. Equation~(2) yields the visibility 
\begin{equation}
V_{\rm co}=\frac{2|a_A|^2|a_B|^2}{|a_A|^2|a_A|^2+|a_B|^2|a_B|^2+2|a_A|^2|a_B|^2}.
\end{equation}
Here $|a_A|^2$  $(|a_B|^2)$  is the classical (constant) intensity of the wave arising from place $A$  ($B$) reaching both $x_1$ and $x_2$ [It appears in both expressions of Eq.~(3)]. The maximum possible value is $V_{\rm co}=\frac{1}{2}$, when $|a_A|^2=|a_B|^2$, in contrast to $V_{\rm qo}=1$ from Eq.~(1).

Now we take into account that the radiation consists of quantum wavepackets. Conceptually, the general procedure in going from the classical to the quantum domain is to re-interpret the classical continuous intensities as probabilities of discrete quantum events. Thus in the quantum domain $|a_A|^2$ is proportional to the probability that wavepacket A (i.e.,    the one coming from place $A$) will give rise to a count either in detector D1 or in detector D2 (even if the arms from B are blocked), and the same holds analogously for $|a_B|^2$.

Accordingly, $|a_A|^2|a_B|^2$ is proportional to the probability that both wavepacket A and wavepacket B will give rise to a count somewhere, these events being independent of each other. In line with this, $|a_A|^2|a_A|^2$ would be proportional to the probability that wavepacket A will give rise to two counts simultaneously (within an arbitrarily short time interval). Classically this would be possible, but owing to the one-quantum nature of the wavepacket it is not. In other words, coincidence counts can only be brought about when both the one-photon packet which comes from A (through $|a_A|^2$) and that which comes from B (through $|a_B|^2$) contribute.  --  Note that all this is independent of whether it was  packet A or packet B that contributed to a count in a particular counter.

Therefore the term $|a_A|^2|a_A|^2$ as well as $|a_B|^2|a_B|^2$ have to be eliminated from Eq.~(4), which then yields a visibility of $V_{\rm qm}=1$, and this coincides with $V_{\rm qo}=1$ of Eq.~(1).

Thus, the situation considered leads to a simple conversion rule for converting the formulas of classical optics into those of quantum optics:  
\begin{quote}
\emph{
In the formulas of classical optics eliminate all those terms that contain products of intensities of waves coming from one and the same place in the source.}
\end{quote}
Further experiments where this rule can easily be seen to apply are: [7] \{Eq. $(8)\mapsto (30)\}$, [8] \{Eq. $(67)\mapsto(72)\}$, [9] \{Eq. $(4)\mapsto(4.33)\}$, 
[10] \{Eq. $(4)\mapsto(41)\}$. The first (classical) formula goes over to the second (quantum) formula by the conversion rule: eliminate $I_1^2, I_2^2$ or $\langle I_1^2\rangle$ or $I_I^2,I_{II}^2$ or $M^2, N^2$ in the respective papers.
\vspace{10pt}

\noindent
{\textbf{3~~Overlap at beam splitters}}
\smallskip

\noindent
Next we consider the paper [11] by Kwiat et al., which is concerned with the coincidence counting rate $P$ of two detectors measured under various polarizations and phase differences of the signal and idler photons. The simplified setup is outlined in Fig.~2. A signal photon $s$ and an idler photon $i$, both with horizontal
\begin{figure}[h]
\begin{center}
\includegraphics[width=0.85\textwidth]{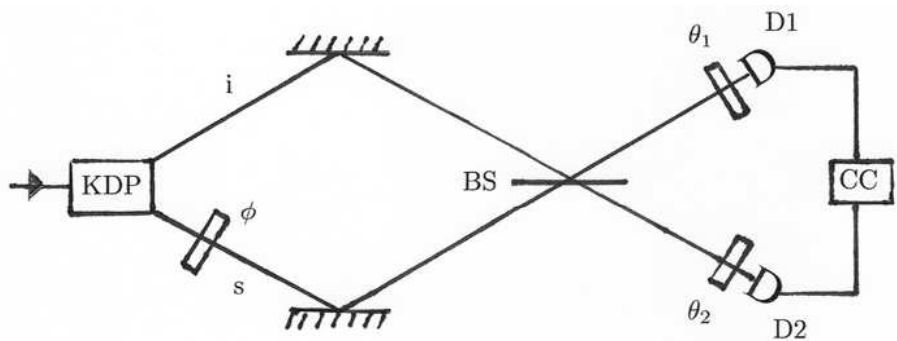}
\caption{Geometry of the interference experiment [11].}
\end{center}
\end{figure}
\noindent
linear polarization, are emitted from the
crystal KDP in different directions, reflected at two mirrors, unified at the 50:50 beam splitter (BS) and led into two detectors D1 and D2, which are connected with a coincidence counter (CC). The setup is symmetrical with respect to the beam splitter, mirrors, and detectors. In the way of the signal photon, a halfway plate is placed, which rotates the plane of polarization by $\phi$. In front of the detectors, linear polarizers can be placed at angles $\theta_1$ and $\theta_2$, respectively, to the horizontal. Each arm contains only one photon at a time.

We consider the coincidence rate $P$ between the two detectors as a function of $\phi$, $\theta_1$, and $\theta_2$. $P$ is proportional to the joint probability considered in Section~2. The experimental results can be described by the quantum-optics formulas in Eqs.~(8), (A4), (13), (A5) in [11]:
\vspace{-5pt}
\begin{equation}
\hspace{1pt}(\rm a)\hspace{65pt} P_{\rm qoa}=\frac{1}{2}\sin^2\phi\hspace{65pt}{\textrm{ without  polarizers}},
\end{equation}

\vspace{-10pt}
\begin{equation}
\hspace{1pt}(\rm b)\hspace{65pt} P_{\rm qob}=\frac{1}{4}\sin^2\phi\hspace{65pt}{\textrm{ with one polarizer}},
\end{equation}
\vspace{-7pt}
\begin{equation}
(\rm c)\hspace{45pt} P_{\rm qoc}=\frac{1}{4}\sin^2\phi\,\sin^2(\theta_1-\theta_2)\hspace{15pt}{\textrm{ with two polarizers}}.
\end{equation}
We now want to derive the formulas to which classical optics would lead us in these situations. Let the classical field amplitudes be $A_{\rm s}$ and $A_{\rm i}$, corresponding to $a_A$ and $a_B$ in the notation of Section~2. At the detectors we consider separately the horizontal H and vertical V components. In classical treatment, the probability of a detector click is proportional to the intensity of the radiation at the place of the detector. Coincidence counts are then proportional to the product of the intensities at detector D1  ($I_1^{\rm H}+I_1^{\rm V}$) and detector D2 ($I_2^{\rm H}+I_2^{\rm V}$) after averaging over the relative random phase $\delta$ between signal and idler pulses:
\[
P_{\rm c}\propto\overline{(I_1^{\rm H}+I_1^{\rm V})(I_2^{\rm H}+I_2^{\rm V})}^{\delta}. 
\]
The intensity at detector D1, say, is just the sum ($I_1^{\rm H}+I_1^{\rm V}$) because there are no interference terms between the horizontal and vertical components. 
\vspace{10pt}

\noindent
Case (a): \\
Only a half-way plate and no polarizers.\\
The horizontal component of the wave at detector D1 is then
\begin{equation}
A_1^{\rm H}=\frac{1}{\sqrt2}A_{\rm s}e^{\rmi(\omega_{\rm s}t+{\bf k}_{\rm s}\bf r_{\rm s} +\delta)}\cos\phi 
+\frac{1}{\sqrt2}A_{\rm i}e^{\rmi(\omega_{\rm i}t+{\bf k}_{\rm i}{\bf r}_{\rm i} +\pi/2)}.
\end{equation}
The factor $\,\exp(\rmi\pi/2)=\rmi\,$ comes from the reflection of the idler wave at the symmetric beam splitter, the factor $1/\sqrt2$ from the splitting in two equal parts, and the factor $\cos\phi$ from the half-way plate in the way of the signal wave. We may omit the terms $\rmi(\omega t+\bf{kr})$ in the exponentials because they can be absorbed in the relative random phase $\delta$. Thus Eq.~(8) simplifies to
\[
A_1^{\rm H}=\frac{1}{\sqrt2}A_{\rm s}\cos\phi\,e^{\rmi\delta}+\frac{\rmi}{\sqrt2}A_{\rm i}.
\]
Analogously, the vertical component of the wave at detector D1 can be written as
\[
A_1^{\rm V}=\frac{1}{\sqrt2}A_{\rm s}\sin\phi \,e^{\rmi\delta}.
\]
In the same manner, at detector D2 we have
\[
A_2^{\rm H}=\frac{\rmi}{\sqrt2}A_{\rm s}\cos\phi\,e^{\rmi\delta}+\frac{1}{\sqrt2}A_{\rm i},\hspace{30pt} A_2^{\rm V}=\frac{\rmi}{\sqrt2}A_{\rm s}\sin\phi \,e^{\rmi\delta}. 
\]
Thus the total intensity at detector D1 becomes
\[
I_1=I_1^{\rm H}+I_1^{\rm V}=|A_1^{\rm H}|^2+|A_1^{\rm V}|^2=\frac{1}{2}\left(A_{\rm s}^2+A_{\rmi}^2+2A_{\rm s}A_{\rmi}\cos\phi\sin\delta\right)
\]
and at detector D2,
\[
I_2=I_2^{\rm H}+I_2^{\rm V}= |A_2^{\rm H}|^2+|A_2^{\rm V}|^2=\frac{1}{2}\left(A_{\rm s}^2+A_{\rmi}^2-2A_{\rm s}A_{\rmi}\cos\phi\sin\delta\right).
\]
Then
\[
I_1I_2=\frac{1}{4}\left((A_{\rm s}^2+A_{\rmi}^2)^2 -4A_{\rm s}^2A_{\rmi}^2\cos^2\phi\,\sin^2\delta\right),
\]
and after averaging over $\delta$ ($\overline{\sin^2\delta}=\frac{1}{2}$), and with $A_{\rm s}^2=I_{\rm s}$ etc., we arrive at
\begin{equation}
P_{\rm co}\propto \frac{1}{4}(I_{\rm s}^2+I_{\rmi}^2) + \frac{1}{2}I_{\rm s}I_{\rmi}\sin^2\phi.
\end{equation}
Applying the conversion rule results in $P_{\rm qm}\propto \frac{1}{2}I_{\rm s}I_{\rmi}\sin^2\phi$, which, except for the unessential proportionality factor $I_{\rm s}I_{\rmi}$, coincides with the quantum optical value in Eq.~(5).

\vspace{10pt}

\noindent
Case (b): \\Half-way plate plus one polarizer.\\
With polarizer angle $\theta_1$ in front of detector D1 classical optics yields

\medskip
$P_{\rm co}\propto \overline{I_1^{\theta_1}(I_2^{\rm H}+I_2^{\rm V})}^{\delta}$

\medskip
$I_1^{\theta_1}=|A_1^{\theta_1}|^2$

\medskip
$A_1^{\theta_1}=\frac{\rmi}{\sqrt2}A_{\rm s}\cos(\theta_1-\phi)+
\frac{1}{\sqrt2}A_{\rmi}\cos\theta_1\,e^{\rmi\delta}$

\medskip
$A_2^{\rm H}=\frac{1}{\sqrt2}A_{\rm s}\cos\phi+\frac{\rmi}{\sqrt2}A_{\rmi}e^{\rmi\delta}$

\medskip
$A_2^V=\frac{1}{\sqrt2}A_{\rm s}\sin\phi$,
\medskip
\newline
and after some calculation, one arrives at
\begin{equation}
P_{\rm co}\propto\frac{1}{4}I_{\rm s}^2\cos^2(\theta_1-\phi)+\frac{1}{4}I_{\rm i}^2\cos^2\theta_1+\frac{1}{4}I_{\rm s}I_{\rmi}\sin^2\phi.   
\end{equation}
Application of  the conversion rule results in
$
P_{\rm qm}=\frac{1}{4}I_{\rm s}I_{\rmi}\sin^2\phi
$, 
which is essentially Eq.~(6) and gives one half the value in Eq.~(5).
\vspace{10pt}

\noindent
Case (c): \\ Half-way plate plus two polarizers.\\
With the polarizer angles $\theta_1$ and $\theta_2$ classical optics yields

\medskip
$P_{\rm co}\propto \overline{|A_1|^2|A_2|^2}^{\delta}$,\\
\vspace{3pt}
\noindent
where

$A_1=\frac{1}{\sqrt2}A_{\rm s}\cos(\theta_1-\phi)+
\frac{\rmi}{\sqrt2}A_{\rmi}\cos\theta_1\,e^{\rmi\delta}$

\smallskip
$A_2=\frac{\rmi}{\sqrt2}A_{\rm s}\cos(\theta_2-\phi)+
\frac{1}{\sqrt2}A_{\rmi}\cos\theta_2\,e^{\rmi\delta}$
\medskip
\newline
and this finally results in
\[
P_{\rm co}\propto\frac{1}{4}I_{\rm s}^2\cos^2(\theta_1-\phi)\cos^2(\theta_2-\phi) +\frac{1}{4}I_{\rmi}^2\cos^2\theta_1\cos^2\theta_2 
\]
\begin{equation}
+\frac{1}{4}I_{\rm s}I_{\rmi}\sin^2\phi\,\sin^2(\theta_1-\theta_2).   
\end{equation}
Applying the conversion rule leaves only the second line of Eq.~(11), which essentially coincides with Eq.~(7).

\vspace{5pt}
Another, more complicated experiment, where the conversion rule by eliminating $\langle I_{\rm s}^2\rangle, \langle I_{\rmi}^2\rangle$, converts the classical formulas into those of quantum optics is [12] \{Eq. (14)$\mapsto$(5a)\}.
\vspace{10pt}

\noindent
{\textbf{4~~No overlap}}
\smallskip

\noindent
Finally, we consider several aspects of an experiment proposed by Franson [13]:
\vspace{5pt}

\hspace{5pt}(A) Its experimental realization by Kwiat \emph{et al.} in the case of narrow coincidence

\hspace{25pt}time windows [14].

\hspace{5pt}(B) Experimental realizations in the case of wide coincidence windows.

\hspace{5pt}(C) A classical calculation for case (A).
\vspace{1pt}
\begin{figure}[h]
\begin{center}
\includegraphics[width=0.45\textwidth]{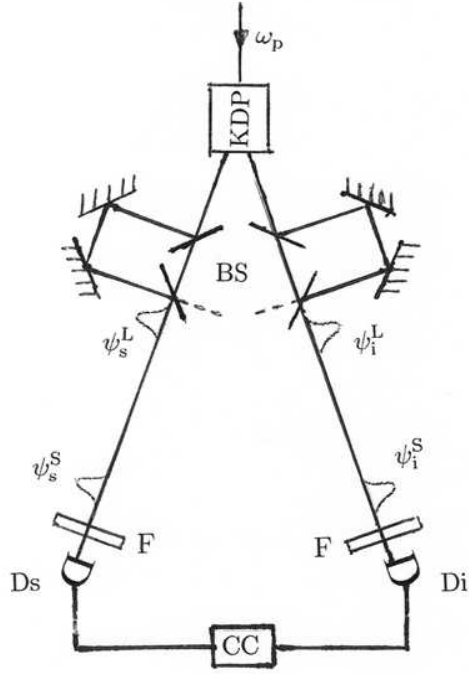}
\caption{Outline of the experimental setup in [14].}
\end{center}
\end{figure}
\vspace{10pt}

\noindent
Case (A):\\
Narrow coincidence window.\\
The principal scheme of the experimental setup [14] is shown in Fig.~3.
The signal and idler photon wavepackets are entangled because they satisfy the conditions
\begin{equation}
t_{\rm s}=t_{\rmi}
\end{equation}
\begin{equation}\bf k_{\rm s}+\bf k_{\rmi}=\bf k_{\rm p}
\end{equation}
\begin{equation}\omega_{\rm s}+\omega_{\rmi}=\omega_{\rm p},
\end{equation}

\medskip

\noindent
where the indices $s, i$, and $p$ refer to the signal photon, the idler photon, and the pump-beam photon, respectively. Each downconverted photon has a substantial bandwidth, but the sum of their frequencies is fixed to within the pump bandwidth, which is negligible in the experiment. The pairs were selected such that $\omega_{\rm s}$ and $\omega_{\rmi}$ are centered at $\omega_{\rm p}/2$. Thus, with $\lambda_{\rm p}$ = 351 nm, $\lambda_{\rm s}$ and $\lambda_{\rmi}$ were each centered at $\lambda$ = 702 nm.

As shown in Fig.~3, each photon packet passes a Mach-Zehnder-like interferometer, where it is split in two equal parts. One part goes the long way and the other the short way, and then the two are recombined. The difference $\Delta L$ between the long and the short way was about 63 cm, the (coherence) length $\sigma_x$ of the signal and idler packets was less than about 36 $\mu$m, and the coincidence time window was $\tau=1.4$ ns, corresponding to a traversed distance of $\delta l=c\tau=44$ cm. Thus $\delta l<\Delta L$, and this means a short coincidence window.
The arrangement is the same for both photons. They are then registered in the detectors, which are assumed to have 100\% efficiency. Filters F of $\Delta\lambda$ = 10 nm bandwidth around $\lambda$ = 702 nm were placed in front of the detectors.

As already stated, there are no definite phase relations between signal and idler photons, and, after leaving the crystal, the two photons never meet again. Yet, as predicted by Franson, based on quantum optics, interference fringes with visibility 1 are observed in the coincidence rates between detectors Ds and Di, when $\Delta L$ is slowly changed.

The explanation based on quantum wavepackets is as follows. In order to construct the wave function of the system of the entangled photons, we start with a product
\begin{equation}
\psius\psiui=(\psiss+\psisl)(\psiis+\psiil)=\psiss\psiis+\psiss\psiil+\psisl\psiis+\psisl\psiil,
\end{equation}
where $\psiss$ is the short-way part, and $\psisl$ the long-way part of the signal wavepacket, $\psius$, and the same for the idler packet, $\psiui$.  Explicitly we write:
\begin{equation}
\psiss({\bf x}_{\rm s},t)=\int_{\bf k_{\rm s}}
\psit({\bf k_{\rm s}})
e^{\rm i({\bf k_{\rm s}\bf x_{\rm s}}-\omega_{\rm s}\it t)}\rmd^3 k_{\rm s}. \end{equation}
The normalized function $\psit(\bf k)$ determines the shape (including the width $\sigma_x$) of the unsplit signal and idler packets as well as of their short-way and the long-way parts. To simplify writing, we write $k$ instead of $\bf k$, and $x$ instead of $\bf x$. $x$ is then a point on the (optical) path counted from the source (the crystal), and $k$ is the component of $\bf k$ in the direction of the path.

Similarly, we then write:
\begin{equation}
\psisl(x_{\rm s},t)=\int_{k_{\rm s}}
\psit({k_{\rm s}})
e^{\rmi({k_{\rm s}(x_{\rm s}-\Delta \it L)}-\omega_{\rm s}\it t)}\rmd k_{\rm s} 
\end{equation}
because we can assume that $\psisl$ has the same shape as $\psiss$, but is displaced by $\Delta L$. Replacing the indices $s$ in Eqs.~(16) and (17) by $i$ gives us $\psiil(\ixi,t)$ and $\psiis(\ixi,t)$, and the first product on the right-hand side of Eq.~(15) becomes
\[
\psiss\psiis=\int_{k_{\rm s}}\int_{k_{\rm i}}\
\psit({k_{\rm s}})\psit({k_{\rm i}})
e^{\rmi(k_{\rm s}x_{\rm s}-\omega_{\rm s}\it t +k_{\rm i} x_{\rm i} -\omega_{\rm i} t)}\rmd k_{\rm s} \rmd k_{\rm i}.  
\]
However, the condition in Eq.~(13) requires the insertion of $\delta(\ki-(k_{\rm p}-\ks))$ [or $\delta(\ks-(k_{\rm p}-\ki)$)] under the integral, which then can no longer be written as a product of a signal and an idler packet, but turns into
\begin{equation}
\psi_{\rm si}^{\rm SS}=e^{\rmi(k_{\rm p} \ixi-\omega_{\rm p} t)}\int\psit(\ks)\psit(k_{\rm p}-\ks)e^{\rmi \ks(\ixs-\ixi)}\rmd k_{\rm s}
\end{equation} 
for any values of $\ixi$ and $\ixs$. Applying the same procedure to the product $\psisl \psiil$ and comparing the result with Eq.~(18) we obtain:
\begin{equation}
\psi_{\rm si}^{\rm LL}=e^{-\rmi k_{\rm p}\Delta L}\psi_{\rm si}^{\rm SS}.
\end{equation}
In the same way, we obtain:
\begin{equation}
\psi_{\rm si}^{\rm SL}=e^{\rmi(k_{\rm p}(\ixi-\Delta L)-\omega_{\rm p} t)}\int\psit(k_{\rm s})\psit(k_{\rm p}-k_{\rm s})e^{\rmi k_{\rm s}(\ixs-\ixi+\Delta L)}\rmd k_{\rm s},
\end{equation}
and $\psi_{\rm si}^{\rm LS}$ is equal to  $\psi_{\rm si}^{\rm SL}$ of Eq.~(20), except that the indices $i$ and $s$ are interchanged.
\vspace{2pt}

Now we apply these formulas to the coincidence probability measured in the experiment [13]. Both detectors were to click at the same time within a small registering time window $\tau\ll\Delta T=\Delta L/c$, and both detectors lie at the same distance from the source, i.e., $\ixs$ = $\ixi$. In this case it is 
\begin{equation}
\psi_{\rm si}^{\rm SL}= \psi_{\rm si}^{\rm LS}=0
\end{equation}
because the integrand in Eq.~(20) is a product of two functions, one of which, $\exp( \rmi k \Delta L)$, oscillates rapidly as a function of $k$, while the other, $\psit(k)\psit(k_{\rm p}-k)$, is a relatively smooth function of $k$. The integral is the closer to zero the larger $\Delta_k\Delta L$, where $\Delta_k$ is the width of the smooth function. In fact $\Delta_k \Delta L \gg 1$ in the considered experiment. With the relation $\Delta_k\Delta_x\approx 1/2$ the condition $\Delta_k\Delta L\gg 1$ can be written as $\Delta L/\Delta_x \gg 1$. As signal and idler packet overlap in $k$ space the width $\Delta_k$ of the product function is comparable to the width $\sigma_k$ of the single parts $\psi_{\rm s}^{\rm S},\,\psi_{\rm s}^{\rm L}$ etc. Thus $\Delta_x\approx\sigma_x$ and $\Delta L/\Delta_x\gg 1$ implies $\Delta L/\sigma_x\gg 1$.
This means that the widths $\sigma_x$ of the parts $\psi_{\rm s}^{\rm S},\,\psi_{\rm s}^{\rm L}$ etc. are small compared with $\Delta L$, which is another way to see that no coincidences between the short-way part of one packet and the long-way part of the other are possible.

Then Eqs.~(19) and (21) yield the coincidence probability
\begin{equation}
C^{(\rm A)}_{\rm qm}=|\psi_{\rm si}|^2=|\psi_{\rm si}^{\rm SS}+\psi_{\rm si}^{\rm LL}|^2=|\psi_{\rm si}^{\rm SS}|^2\,|1-e^{-\rmi k_{\rm p}\Delta L}|^2=|\psi_{\rm si}^{\rm SS}|^2\,2\,(1+\cos k_{\rm p} \Delta L),
\end{equation}
and this means visibility $V_{\rm qm}=1$, the same as predicted by quantum optics. The measured value of only $0.8<1$ can be ascribed to imperfections in the experimental setup.
\vspace{10pt}

\noindent
Case (B):\\
Wide coincidence window.\\
Both detectors may click within a coincidence time window, $\tau > \Delta T$. Such a case was studied in the experiments [15-17]. In this case, the coincidence probability is the sum of three probabilities, which refer to three different physical situations:

\vspace{5pt}
(B1) Coincidences between the short-way parts of the signal and the idler packet and between the long-way parts of the two packets.
 
(B2) Coincidences between the short-way part of the signal packet and the long-way part of the idler packet.

(B3) Coincidences between the short-way part of the idler packet and the long-way part of the signal packet.
\vspace{7pt}

Case (B1) is equivalent with detectors which have the same narrow coincidence window  and lie at the same distance from the source, $\ixs-\ixi=0$. This is the case already treated in case~(A), where $C_{\rm qm}^{\rm (A)}$ is given by Eq.~(22), which with Eq.~(18) reads
\begin{equation}
C_{\rm qm}^{(\rm B1)}=\left|\int\psit(k_{\rm s})\psit(k_{\rm p}-k_{\rm s})\rmd k_{\rm s}\right|^2 2\,(1+\cos k_{\rm p}\Delta L). 
\end{equation}

Case (B2) is equivalent with detectors which have the same narrow coincidence  window, and detector Ds lies at $x_{\rm s}=x_{\rmi}+\Delta L$ while Di lies at $x_{\rmi}$; that is, $x_{\rm s}-x_{\rmi}=+\Delta L$. In this case, the integral for $\psi_{\rm si}^{\rm SS}$ [Eq.~(18)] contains the factor $\exp(+\rmi k_{\rm s} \Delta L)$, which is now the rapidly oscillating factor. By a reasoning like that after Eq.~(21) the integrals in Eqs.~(18) and (19) now result in 
\[
\psi_{\rm si}^{\rm SS}=\psi_{\rm si}^{\rm LL}=0.
\]
The integral for $\psi_{\rm si}^{\rm SL}$ [Eq.~(20)] contains the factor $\exp(\rmi k_{\rm s}(\ixs-\ixi+\Delta L))$, which with $\ixs-\ixi=+\Delta L$ becomes
the rapidly oscillating factor and leads to
\[
\psi_{\rm si}^{\rm SL}=0.
\]
The integral for $\psi_{\rm si}^{\rm LS}$ contains the factor $\exp(\rmi k_{\rm s}(\ixs-\ixi-\Delta L))$, which with $\ixs-\ixi=+\Delta L$ becomes 1 and leads to 
\[
|\psi_{\rm si}^{\rm LS}|^2=\left|\int\psit(k_{\rmi})\psit(k_{\rm p}-k_{\rmi})\rmd k_{\rmi}\right|^2\equiv\left|\int\psit(k_{\rm s})\psit(k_{\rm p}-k_{\rm s})\rmd k_{\rm s}\right|^2.
\]
Thus
\begin{equation}
C_{\rm qm}^{(\rm B2)}=|\psi_{\rm si}^{\rm LS}|^2=\left|\int \psit(k_s)\psit(k_{\rm p}-k_s)\rmd k_s\right|^2.
\end{equation}

Case (B3) means that $x_{\rm s}-x_{\rmi}=-\Delta L$, and repeating the steps that led us in (B2) to Eq.~(24) results in 
\begin{equation}
C_{\rm qm}^{(\rm B3)}= C_{\rm qm}^{(\rm B2)},
\end{equation}
which is as it should, due to the signal/idler symmetry.
With Eqs.~(23), (24), and (25) the total coincidence probabilitiy in case (B) is
\begin{equation}
C^{(\rm B)}_{\rm qm}=C_{\rm qm}^{(\rm B1)}+C_{\rm qm}^{(\rm B2)}+C_{\rm qm}^{(\rm B3)}=\left|\psi_{\rm si}^{\rm LS}\right|^2 4 (1+\frac{1}{2}\cos k_{\rm p}\Delta L). 
\end{equation}
This means visibility $V_{\rm qm}=\frac{1}{2}$, which is confirmed in the above-mentioned experiments.

Cases (A) and (B) were realized in the experiment [18] and can be treated in the same way as the experiment in [14], though the paths of the signal and idler photons in [18] were not spatially separated. The visibility obtained was $V=0.87$ in case~(A) and $V=0.46$ in case~(B). 
\vspace{10pt}

\noindent
Case~(C):\\
A classical calculation.\\
It is interesting to compare the quantum case (B1), where $x_{\rm s}=x_{\rm i}$, $V_{\rm qm}$ = 1, with the following classical calculation. A possible classical analog would be a situation where the electromagnetic signal and idler pulses are independent of each other. This might be expressed explicitly by multiplying the signal pulse, say, by $\exp(\rmi\delta)$ with random $\delta$, but such a factor would drop out in the calculations below.

In the classical case, the signal and idler pulses even before measurement lie in rather narrow intervals about $\omega_{\rm s},\, \ks$, and $\omega_{\rm i},\, \ki$, respectively, albeit still permitting sufficiently short packets in $x$ space. Then a statistical ensemble of different runs is considered, where different  runs have different values of $\omega$ and $k$, satisfying, however, the conditions in Eqs.~(12), (13), and (14) in each single run.

In the quantum case, the coincidence rate was obtained by first averaging (integrating over $\ks$ in a rather wide interval) the product of the signal and idler \emph{amplitudes}, meaning entanglement, and then taking the absolute square. In the classical case, waves may be added but never multiplied. Classically only intensities may be multiplied (cf. Section~3). Thus we calculate the classical \emph{intensities} at detectors Ds and Di, multiply the two, and then average the product over $\ks$ and $\ki$. Such a procedure is like that applied to the case of two spin-1/2 particles in a spin-singlet state in [5, Appendix C].

Thus we again start with a product of a signal and an idler pulse,
\begin{equation}
E_{\rm s} E_{\rm i}=(E_{\rm s}^{\rm S}+E_{\rm s}^{\rm L})(E_{\rm i}^{\rm S}+E_{\rm i}^{\rm L})=
E_{\rm s}^{\rm S}E_{\rm i}^{\rm S}
+E_{\rm s}^{\rm S}E_{\rm i}^{\rm L}
+E_{\rm s}^{\rm L}E_{\rm i}^{\rm S}
+E_{\rm s}^{\rm L}E_{\rm i}^{\rm L}
\end{equation}
as in Eq.~(15). Here we write $E_{\rm s}^{\rm S}$ etc. instead of $\psi_{\rm s}^{\rm S}$ etc. in order to emphasize that our wavepackets are now pulses of classical electromagnetic waves. As $k$ and $\omega$ now lie in narrow intervals, we approximate the pulses by integrals like those for $\psi_{\rm s}^{\rm S}(\ixs,t)$ in Eq.~(16), where, however, the range of the integration is so small that we can approximate the integrals by their integrands multipied by $\Delta k$. For example [cf. Eq.~(16)], 
\[
E_{\rm s}^{\rm S}(x_{\rm s},t)=
\psit({k_{\rm s}})
e^{\rm i({k_{\rm s} x_{\rm s}}-\omega_{\rm s}\it t)}\Delta \ks \hspace{50pt} 
\]
\[
E_{\rm s}^{\rm L}(x_{\rm s},t)=
\psit({k_{\rm s}})
e^{\rm i({k_{\rm s} (x_{\rm s}-\Delta L)}-\omega_{\rm s}\it t)}\Delta \ks, \hspace{50pt}
\]
and analogous formulas for $E_{\rm i}^{\rm S}$ and $E_{\rm i}^{\rm L}$. Then the product in Eq.~(27), with $\ki=k_{\rm p}-\ks$ from Eq.~(13), turns into
\[
E_{\rm s}E_{\rm i}=E_{\rm s}^{\rm S}E_{\rm i}^{\rm S}\left(1+e^{-\rmi(k_{\rm p}-\ks)\Delta L}+e^{-\rmi\ks\Delta L}+e^{-\rmi k_{\rm p}\Delta L}\right)\Delta\ks\Delta\ki.\hspace{50pt}
\]
In averaging $\overline{\left|E_{\rm s}E_{\rm i}\right|^2}\; (\Delta k\rightarrow \rmd k$ and integrating over a wide interval) all those of the 16 terms which contain factors like $\exp(\pm \rmi\ks\Delta L)$ average to zero, following the reasoning after Eq.~(21). The surviving terms are
\begin{equation}
\overline{\left|E_{\rm s}E_{\rm i}\right|^2}
=\overline{\left|E_{\rm s}^{\rm S}E_{\rm i}^{\rm S}\right|^2}
\left(4
+e^{+\rmi k_{\rm p}\Delta L}
+e^{-\rmi k_{\rm p}\Delta L}\right)=\overline{\left|E_{\rm s}E_{\rm i}\right|^2}4(1+\frac{1}{2}\cos k_{\rm p}\Delta L),
\end{equation}
which means visibility $V_{\rm co}=\frac{1}{2}$, that is half the quantum value  in Eq.~(22) in the situation considered.
\vspace{5pt}

In essence, the term $k_{\rm p}\Delta L=(\ks+\ki)\Delta L$ arises from  the product of the signal and idler waves and also from the product of the classical intensities. The particular value $V_{\rm qm}=1$ arises from averaging over the product of the amplitudes, rather than over the product of the intensities, that is, from entanglement.
\vspace{10pt}

\noindent
{\textbf{References and notes}}

\begin{enumerate}
\renewcommand{\labelenumi}{[\arabic{enumi}]}

\item A. Einstein, ``\"Uber einen die Erzeugung und Verwandlung des Lichtes betreffenden heuristischen Gesichtspunkt,'' Ann. Physik (Leipzig) {\bf17}, 132-148 (1905). English translation by A.B. Arons and M.B. Peppard, ``Einstein's proposal of the photon concept - a translation of the Annalen der Physik paper of 1905,'' Am. J. Phys. {\bf33}, 367-374 (1965).

\item A. Jabs, ``A conjecture concerning determinism and phases in quantum mechanics,'' arXiv:1204.0614.

\item A. Jabs, ``Picturing wavepacket reduction,'' arXiv:1305.5119.

\item A. Jabs, ``An interpretation of the formalism of quantum mechanics in terms of epistemological realism,'' arXiv:1212.4687 (Br. J. Philos. Sci. {\bf43}, 405-421 (1992)).

\item A. Jabs, ``Quantum mechanics in terms of realism,'' arXiv:quant-ph/9606017 (Physics Essays {\bf9}, 35-95, 354 (1996)).

\item R. Ghosh, and L. Mandel, ``Observation of Nonclassical Effects in the Interference of Two Photons,'' Phys. Rev. Lett. {\bf59}, 1903-1905 (1987). 

\item L. Mandel, ``Photon interference and correlation effects produced by independen quantum sources,'' Phys. Rev. A {\bf28} (2), 929-943 (1983). 

\item L. Mandel, ``Non-Classical States of the Electromagnetic Field,'' Physica Scripta {\bf T12}, 34-42 (1986). 

\item H. Paul, ``Interference between independent photons,'' Rev. Mod. Phys. {\bf58} (1), 209-231 (1986). 

\item G. Richter, ``\"Uber Interferenzen und Intensit\"atskorrelationen bei Lichtstrahlen sehr geringer Intensit\"at,'' Ann. Physik (Leipzig) {\bf45} (8), 564-577 (1988). 

\item P. G. Kwiat, A. M. Steinberg, and R. Y. Chiao, ``Observation of a ``quantum eraser'': A revival of coherence in a two-photon interference experiment,'' Phys. Rev. A {\bf45} (11), 7729-7739 (1992). 

\item Z. Y. Ou, and L. Mandel, ``Violation of Bell's Inequality and Classical Probability in a Two-Photon Correlation Experiment,'' Phys. Rev. Lett. {\bf61} (1), 50-53 (1988). 

\item J. D. Franson, ``Bell inequality for Position and Time,'' Phys. Rev. Lett. {\bf62} (19), 2205-2208 (1989). 

\item P. G. Kwiat, A. M. Steinberg, and R. Y. Chiao, ``High-visibility interference in a Bell-inequality experiment for energy and time,'' Phys. Rev. A {\bf47}, R2472-R2475 (1993). 

\item P. G. Kwiat, W. A. Vareka, C. K. Hong, H. Nathel, and R. Y. Chiao,  ``Correlated two-photon interference in a dual-beam Michelson interferometer,'' Phys. Rev A {\bf41} (5), 2910-2913 (1990). 

\item Z. Y. Ou, X. Y. Zou, L. J. Wang, and L. Mandel, ``Observation of Nonlocal Interference in Separated Photon Channels,'' Phys. Rev. Lett. {\bf65} (3), 321-324 (1990). 

\item J. D. Franson, ``Two-photon interferometry over large distances,'' Phys. Rev. A {\bf44} (7), 4552-4555 (1991). 

\item J. Brendel, E. Mohler, and W. Martienssen, ``Time-Resolved Dual-Beam Two-Photon Interferences with High Visibility,'' Phys. Rev. Lett. {\bf66} (9), 1142-1145 (1991). 

\end{enumerate}
\bigskip
\hspace{5cm}
------------------------------
\end{document}